# A Design Guideline to Overcome Web Accessibility Issues Challenged by Visually Impaired Community in Sri Lanka

N Wedasinghe[1#], NT Sirisoma[2], and APR Wickramarachchi[3]

[1,2] *General Sir John Kotelawala Defence University*

[3] *University of Kalaniya, Faculty of Science*

[#]<nirosha@kdu.ac.lk>

*Abstract* - Visual impaired community are the one of the hindrance group of accessing web content access in the world. The obstacles encountered by this community in their current practices and to develop a best practice guidelines to overcome the digital divide in Sri Lanka becomes gap filling of this domain. Preliminary survey indicated that five main problems including 1.) Access limited by the impairment, 2.) Usability issues due to lack of designing, 3.) Unavailability of visually impaired-friendly applications, 4.) Lack of communication, 5.) Web navigation issues are the most dominant pertaining issues. To overcome those issues, solutions are tested and validated with using Design Science approach. Purposive sampling method used and Interviews and questionnaires are used to extracte data. Content analysis also used to derive the result. Results are further validated with using expert opinion. Result Indicate that significant factors need to be incorporate are, ensuring a keyboard-friendly websites, easily accessibility and support with semantic annotation by adding alternative text for images. Furthermore, use headers to structure the content correctly, design all forms to support accessibility in mind including Content developing and designing, navigation, the best colour combination, Pre-recorded video with the audio facilities, braille support in web, the designing option has no significant impact on visually impaired web users. Introducing a rating widget option to a website identifies the level of accessibility features availability facilitates, thereby overcoming the disability digital divide. The results further conclude that a significant difference exists in websites, with and without the involvement of the visually impaired community. Semantic web and semantic annotations of the context of page elements, content serialisation*,* and navigation by special keyboard commands are also highly influencing the effective use of the web and increase the satisfaction level in the website accessing process.

## I. INTRODUCTION

According to the Ceylon medical journal (Murthy et al., 2018), 285 million people are visually impaired globally from which 39 million are blind. A survey conducted by the Ministry of Health in Sri Lanka (Sri Lankan Ministry of Health Report (2016) found that the occurrence of blindness in Sri Lanka population is 1.7% among the age group above 40. Further, around 9.3 million people in Sri Lanka suffer from some form of vision impairment NPPAB (2020).

One of the biggest issues that visually impaired population experience is difficulties in accessing Information Technology based services including access to the internet BOIA (2019) and Hollier (2007). To a certain extent, the advances in information technology made possible to provide access to digital devices and web-enabled services such as online banking, e-government services, online shopping etc. to differently abled community. However, due to various economic and social issues, visually impaired communities in Sri Lanka still confront many difficulties in accessing the web-based services. Analysis reveals that there is no detailed studies are conducted on how such issues are addressed through policies and guidelines to reduce the disability digital divide faced by the visually impaired community in the country. The objective of this paper is to propose a set of guidelines to overcome the disability digital divide among the visually impaired community in Sri Lanka.

## II. LITERATURE REVIEW

Visual Impairment (VI) is a significant type of impairment. Various scholars have defined it visually to be inclusive of blindness, while others exclude blindness in their definitions. According to the WHO (2019), The International Classification of Diseases 11 (2018) classifies VI into different groups, remoteness, near presenting, and distance vision impairment: "Blindness" – presenting visual acuity not as good as than 3/60, "Moderate" – presenting visual acuity not as good as than 6/18, "Mild" – presenting visual acuity not as good as than 6/12, "Severe" – presenting visual acuity not as good as than 6/60.

*A. Obstacles in Information Inaccessibility*

The W3C (2016) briefly introduces five categories of visual impairment that impact in web use. They are Visual acuity (clarity), Light sensitivity, Contrast sensitivity, Field of vision, and Colour vision. In the context of information accessibility, visually impaired individuals are more disadvantaged than the other disabled categories and non-disabled categories. It creates a gap of accessing information which causes the digital divide.

*B. Web Accessibility Issues*

Raufi (2015)'s study on web accessibility revealed that even though web pages act following Web Content Accessibility Guidelines (WCAG), visually impaired users still face route-finding problems (Navigational) (Power et al., 2012). Ferati et al. (2014) concludes that for blind and visually impaired users, software solutions are not a priori guarantee of digital content accessibility; hence many aspects including the cultural ones need to be considered (Ferati et al., 2014). Screen reader for the blind is sequentially navigation on the web through nature, although the Internet offers parallel and increasingly non-sequential content.

Web accessibility problems affecting blind users identified by researchers Leporini and Paterno (2008) include:

**Lack of page context:** Users tend to lose the overall navigational context when visiting and reading minor portions of text within the page. Overcoming this obstacle is possible through the semantic web and semantic annotations of the context of page elements (Fernandes, 2006; Semaan, 2013).

**Information surplus due to extreme sequential reading:** Static portions of the web documents are often overloaded with links, frames, headers, menus, and footer elements that obstruct the reading process—hence, a large amount of unnecessary data has to be excluded for reaching the desired information. A possible answer that addresses this issue is the concurrent speech method. This method allows blind individuals to find information more speedily (Guerreiro and Gonçalves, 2014; Ahmed et al., 2012). The following design aspects are found obtrusive when considering the visually impaired operators.

**Content serialisation:** Typically, all content blocks are presented (serialised) in a sequential order without considering the design aspects and relative positioning within hyper-documents.

**Navigation by special keyboard commands:** Visually impaired users prefer conducting their navigation through specialised keys, which ensure them quick access to required information. Thus, providing keyboard functionalities is required either through a specific feature in reading software or as a unique tag within hyper-document.

**The difference in information conveyance between visual layouts and those afforded by aural perception:** Often, for users with no sight issues, when secondary information is given, they can identify it immediately (left or right menu bars, special headers, etc.). Such information must be presented conveniently for visually impaired individuals as well. One such approach, especially for graphs and tables, is seen in MultiVis Project (Kildal and Brewster, 2007) where haptic and non-speech sounds are used for making data visualisations (Silva,2016).

*C. Usability and Accessibility Issues of Websites in Sri Lanka*

According to Gopinath et al. (2016), web analysis of Sri Lankan Government websites, usability and accessibility standards survey results indicate that websites must ensure that the e-government provides not only better service to the public but also a convenient service to the vision impaired community of the country. They further stated that it is legally and ethically essential to building e-government sites accessible by visually impaired users. They have already done a survey and identify Most of the Sri Lankan e-government websites do not strictly follow the WCAG 2.0 rules. With using "WAWE" accessibility software 45 government official web sites tested and found 45 are error – free but more than 90 percent of them are faced with inconvenience or inaccessibility for persons with different disabilities. They have already mention that necessity of WCAG 2.0 success criterion (Level AA and Level AAA) .

Furthermore users who use assistive technologies such as JAWS, NVDA, and WindowsEyes will often find difficulties in navigating the website, obtaining information, completing a task or performing a search request. Without being able to navigate the web page, it is impossible to perform a task no matter how accessible the website is. (Gopinath, 2016).

Therefore, requirement of proper guidelines to follow becomes an important aspect to the country.

According to the Wedasinghe et. al (2014,2015,2017,2018,) studies and literature revels that when this community accessing Web applications, there are five identified categories of problems associated. They are including: 1.)Access limited by the impairment, 2.)Usability issues due to lack of designing, 3.)Unavailability of visually impaired-friendly applications, 4.) Lack of communication, 5) Web navigation issues.

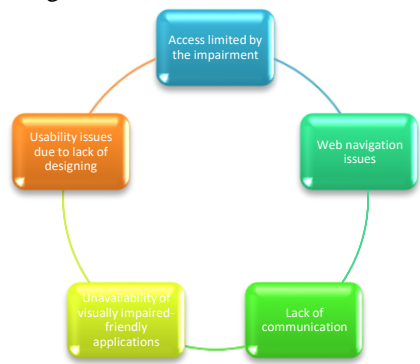

Figure 1: Categories of web accessibility issues

### III. METHODOLOGY

For each of the above category proposed solutions were tested in the five different phases. Based on the result recommendations were proposed to develop the guidelines. Policymakers, web designers and developers could use these recommendations in guidelines preparations, and in web and mobile application development process. The researcher used a design science approach, which is mostly applied for information technology research. In the different phases different research methods used.

*A.Data Collection and Analysis Techniques*

In the First Phase consider on the Access limited by the impairment.In this phase check different checklist to ensure that the identified features are in the existing web sites. Select two different types of websites, one with the above facilities, and other without the facilities.

In the Second phase is consider on the usability issues due to the lack of designing. This activity is the consideration of brail display support and audio device support in the user interface design process to improve usability.

Selecting two different types of websites; One is provided with brail and audio facilities, and the other is without the facilities.

Phase Third is consider about the Waiting to Access Visually Impaired Friendly Version to access the website.To check this Website designers and owners can check the website for compatibility for blind and visually impaired before publishing to the public access to overcome this issue.

In this activity, the experimental method and the following steps were used for validation.A sample website, including visual impaired-friendly options, were added (Text resizing option, Picture description, Voice output option).Users are given an additional option for rating the website by adding a 'rating' option on the site.The feedback of the visually impaired users are evaluated, and conclusions are derived based on the analysis.

Fourth Phase is consider on Lack of Communication issue.This activity includes the visually impaired community in the process of web design and development. This activity is done by developing one website with the involvement of three blind individuals and another three with low vision individuals. The other website is designed, tested, and developed without the participation of the vision-impaired community. Next, these two websites were given to ten blind and low vision individuals to check the satisfaction level. A questionnaire was given to these participants with answers in a Likert scale (No satisfaction - 0, Little satisfaction - 1, Moderate satisfaction - 2, High satisfaction - 3 and Very high satisfaction - 4) to monitor the significant differences in the websites used with the involvement of visually impaired users and without commitment.

In the fifth phase is focuses on the web navigation related issues. This activity was evaluated using five expert opinions in the field of web designing and development— this expertise was selected by using their expertise, especially on web designing and development for visually impaired requirements. Responses were collected using a Google Form, and data were presented using a pie-chart.

This research is an outcome-based research approach, where research focuses on the development and performance of the artefacts with the explicit purpose of improving its functional performance.

## IV. ANALYSIS

The above discussed five Pases research activities analysied and results are present in this section.

**Phase 1 - Access Limited by the Impairment**

The low vision individuals required screen magnifiers, screen reading software and voice recognition software. Therefore, website designers and developers must consider the assistive software compatibility on websites. Following ten compatibility checks tested to ensure that websites facilitate the following features:

- Make sure the website is Keyboard-friendly.

A visually impaired user has no choice of using a mouse. They cannot locate where to click the mouse. Not as mouse access keyboard is sequential, access such as blind user needs to tab key unless reaching the target destination. Therefore, keyboard-friendly interactive websites become useful and possibly allow these users to access the Internet. It will assist in overcoming the web accessibility issues. Therefore, the following compatibility checks were identified as web page developments:1.)Make sure all content is easily accessible and support with Symantec annotation. 2.)Add Alt Text to All Images (Alternative Text for images). 3.)Use headers to structure the content correctly.4.) Design all forms for accessibility.5.)Do not use Tables for whatever except tabular data.6.)Enable resizable text that does not break a website.7.)Avoid automatic media and navigation. 8.)Create content with accessibility in mind. 9.)Use the best colour combination for Visually Impaired Community.

Each point given in the above checklist is validated with two different types of websites. One is with a facility highlighted above, and the other one is without a facility. For this validation well known sri Lankan websites such as "http://sinhala.adaderana.lk/" , "http://www.lankadeepa.lk/" ," http://www.airforce.lk/", "http://topjobs.lk/", "https://www.csdeafblind.lk/"  and spme more popular web sites in Sri Lanka.

A focused group, including ten visual impaired individuals, were selected for validation. A sample is selected with three fully visually impaired, another three with partial visually impaired, and the rest with mild vision difficulties. The participant is surfing the websites and experience the readiness of the websites about the criteria. Next, a five-point Likert scale is used to validate what rating is given for the website with/without the proposed designing features. The Likert scale is divided into following five scales and values starting from 0 to 4 are given; Excellent - 4, Good - 3, Fair - 2, Poor - 1, and Very poor - 0. Finally, values will help to conclude how the proposed design is affected by accessibility. The average of five-point Likert scale values given by totally blind participants with proposed facilities (AVFTB) was calculated using the following formula:

$$AVFTB = \sum_{i=0}^{n} V$$

n= Number of participants, V = five-point Likert scale values given by participants, i=counter

Furthermore, the average of five-point Likert scale values given by totally blind participants without proposed facilities (AVWFTB) was calculated using the following formula:

$$AVWFTB = \sum_{i=0}^{n} V$$

n= Number of participants, V = five-point Likert scale values given by participants, i=counter

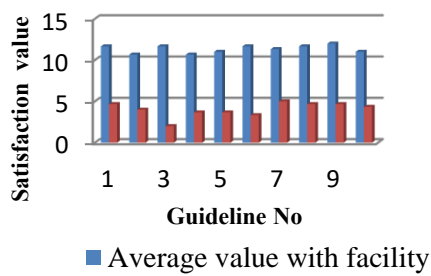

**Figure 2: Summary of satisfaction level on different accessibility functions**

According to the findings in Figure 2 the bar chart clearly shows that features discussed have significant effects on visually impaired individuals. Therefore, it is recommended for web designers and web developers to increase accessibility among the visually impaired community using these discussed features.

## Phase 2 - Usability Issues Due to Lack of Designing

This section considers solving usability related issues. Two usability related developments were identified in website developing, as presented:
- Braille display (Braille Terminal) support
- Audio support

The compatibility of braille display is an important factor that the website designers and developers must consider. To serve braille readers adequately, the web developers and designers must evaluate their sites' compatibility with refreshable braille displays. It is recommended to consider WCAG(2.1) guidelines. Therefore, under the WCAG (2.1) recommendations, web designers and developers must consider the following tips (BOIA, 2019):

- The alt HTML attribute used to provide captions for images and another non-textual context. For example, the HTML code  provides alternative text for a photograph posted in the breakfast in the morning.
- The lang HTML attribute used to specify the natural language of a given web page or passage. For example, the HTML code <html lang="en"> designates that the page is written in English.

Audio support is necessary to consider when the websites contain only a **pre-recorded video.** Web designers and developers must find either an alternative for time-based media or an audio track to provide for visually impaired users.

Three visually impaired individuals were selected, and they were provided with facilities to access a website containing the above features, and others without support. Participants were interviewed to get their feedback on the usability issue. Specific features were monitored, including how the websites that offer brail support and audio support in web usability.

Five experts were interviewed for further validation. According to the findings, it indicated that the brail support is not directly influencing to regular web users, but it directly affects if the content is printing via the web to brail. According to the expert opinion "if the website does not have the brail facility, then it can be converted to the word document and print in the brail without much difficulty.
According to the feedback concludes that audio support is an essential part of web accessibility due to vision difficulty. Participant results showed that 44% of them select a website with audio support, which is convenient for them. Another 37% of them also mentioned that they are satisfied with the audio backing, and the rest of the 26% of the participants mentioned that the website is straightforward to use with audio support.

Participants without the support of audio in the website, and 37 per cent of them showed that "Audio is compulsory; otherwise difficult to understand" and 33 per cent of the participants mentioned that "Difficult me to continue without audio support." The rest of the 30 participants showed that "Very inconvenient without audio support."

Furthermore, five expert validations also conducted, and all experts emphasized that the voice support for audio is essential to consider. Therefore, web developers need to consider providing a pre-recorded video with audio facilities to overcome the web related issues.

## Phase 3 - Waiting to Access Visually Impaired-Friendly Version

As a solution to this issue, propose website designers and owners check websites for compatibility for blind and visually impaired before publishing to the public access. Then vision impaired persons who log in to the site recognized from the opening about the accessibility level of the website rather than a frustrating reading of the entire website. Therefore, the rating widget and accessibility widget installed into a website are more appropriate for visually impaired persons to show the accessibility level. This method is tested to increase the popularity of the website and motivation for other website owners to increase the website's popularity. It is like most popular travel agents adding their customer rates at the top corner of their websites. Adding this option increased the popularity of the website among the visually impaired individuals.

In this activity, add an accessibility compliance-rating badge to a website, following steps tested using the experimental method.

Design a website, including visually impaired-friendly options added with following features including Keyboard support, Text resizing option, Picture description (Alter Text), Brail output option, Voice output option, Download a web content and features a testing tool, Website fully compatible with accessibility features are given with a visually impaired compliance badge can include in an appropriate place on the website. Figure 3 shows the accessibility badge.

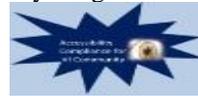

**Figure 3 : Accessibility Badge**

Users are given an option to rate the website (according to the accessibility features). Adding a 'rating' option on the website (Adding a rating widget). This rating is always presented on the website.

This activity validated using a focused group with 24 visually impaired individuals. Ten of them were Completely vision loss, and the other fourteen were with low vision. First, they were given the website to access accessibility features and subsequently given five more websites with low accessibility features.
Then they were provided with a Google feedback form to get their response on accessibility.

Figures 4 and 5 present responses, using area charts.

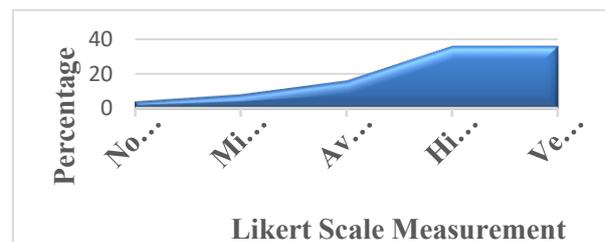

**Figure 4 Response for User feedback on Impact for accessibility badge option**

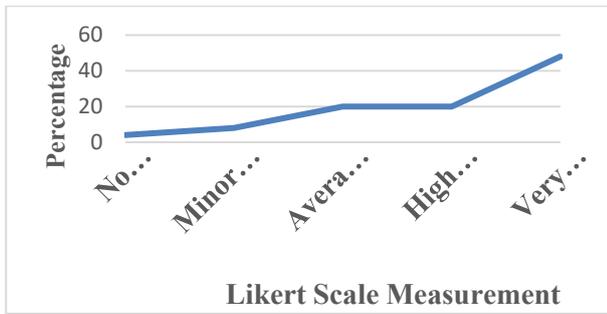

**Figure 5: Response for awareness of web developers on accessibility badge option**

The response indicated that 36% of them agreed that including an accessibility badge is a very high impact, and another 36% of them have noted its high impact. Only 16% indicated the impact as only moderate. Another 8% suggests that there is little impact and the rest of the four indicated no effect.The majority of visually impaired respondents, which includes 72% of the total respondents, identifies that the impact of having an accessibility badge on a website is high. Therefore, according to the result, it becomes an essential factor to implement.Next, the website owners or designers' view whether it is necessary to consider an accessibility badge was checked at the time of designing and developing a website. Out of all respondents, 48% indicated that it is very high, 20% stated that it is high, 20% reported it is moderate, and 08% indicated little, while the rest of 4% indicate no impact. Since the majority of 68% suggested it as necessary, it is concluded that this accessibility badge is an essential and a considerable web accessibility feature that could be used in the website to overcome the disability digital divide issues faced by the vision-impaired community. This was further validated with eight more expertise in web development and two visually impaired individuals who are experts in the web accessibility area. The response further verified that the result is reliable.

### Phase 4 - Lack of Communication

This activity is to consider the effectiveness of including the visually impaired community in the process of web designing and development to overcome the web accessibility issues. Web designing and development are starting from the requirements of the gathering stage, designing and development of the website, followed by implementation of the website. In every step in this process, it is necessary to actively involve the visually impaired individuals to get the real requirements and testing of the website accessibility requirements. They have practical experiences of accessibility issues.

This activity is validated by developing one website with the completely vision loss 03 individuals and others with low vision impairment. Another website was tested designing and developed without the participation of the visually impaired community.

The two developed websites were given to ten fully visually impaired and partially visually impaired individuals to detect the user-friendliness of the website and their satisfaction levels. These participants were given a questionnaire with answers in a Likert scale (No satisfaction - 0, Slight satisfaction - 1, Moderate satisfaction - 2, High satisfaction – 3, and Very high satisfaction - 4), to monitor the significant difference in the websites developed with and without involving visually impaired users.

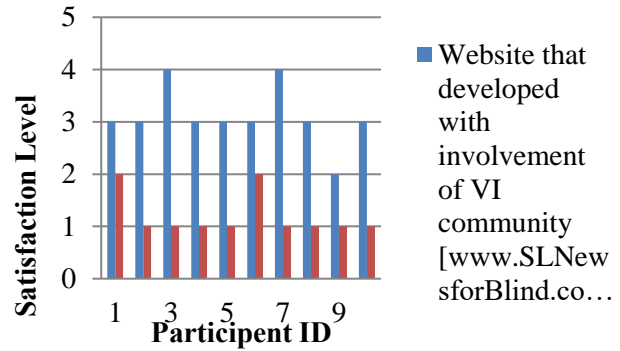

**Figure 6: Evaluating VI involvement and without the involvement of website designing**

According to the responses, 90% of the participants were with high and very high satisfaction levels for the website that developed with the involvement of visually impaired participants. The other news website received light and moderate satisfaction levels. Therefore, it encourages website designers and owners to include visually impaired individuals in the website development process. For testing the website using web accessibility compliance tools, four experts tested and validated the website to improve the reliability of the result.

### Phase 5 - Web Navigation Issues

Navigation on the web through a screen reader for the blind is sequential by nature, although the web offers parallel and increasingly non-sequential content. There are two main problems addressed in this activity. The first one is the **lack of page context.** In this, visually impaired users are navigational context when they are visiting another web page to read the small information portion of the text. Therefore, the overall focus of the content is challenging to keep. Comparatively, with non-visually impaired web users, visually impaired users get more frustrated in designing these web pages. This can overcome through semantic web and semantic annotations of the context of page elements.

The next problem address is **Information overload due to excessive sequential reading.** Static portions of the web documents often overloaded with links, frames, headers, menus, and footer elements that obstruct the reading process. Unnecessary data must omit to identify the desired information to reduce of getting frustrated in this web accessing process.

A possible solution that addresses this issue is, either using concurrent speech methods that allow blind users to find information more quickly or through skimming (i.e., grasping the tip of the information to determine its content quickly).

In addition to the above solution, content **serialization** is also one alternative. In this method, typically, all content blocks presented (serialized) in a sequential order without considering the design aspects and relative positioning within hyper-documents.

Another solution proposed is **navigation by special keyboard commands.** Visually impaired users prefer conducting their navigation through specialized keys. Therefore, ensuring keyboard functionalities is required either through a specific feature in reading software or as a unique tag within hyper-document.

Two websites evaluated in this activity were compared. One website was developed with Symantec web option, content serialization, and navigation through particular function key option. Another website was selected that does not contain the above three features. Then selected a focused group with twenty visually impaired samples, including fully visually impaired and low vision people. After using the websites, a Google form questionnaire was given for them to fill.

The analysis of the responses is shown below, and the following URL inactivate online.

Responses are presented in Figures 7 to Figure 9, using pie charts.

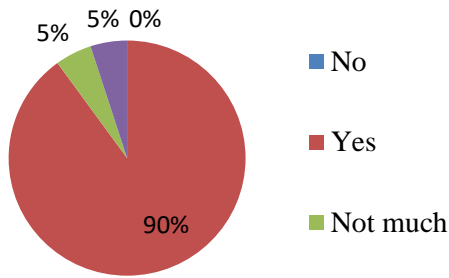

**Figure 7: Responses for navigation issues in website 1**

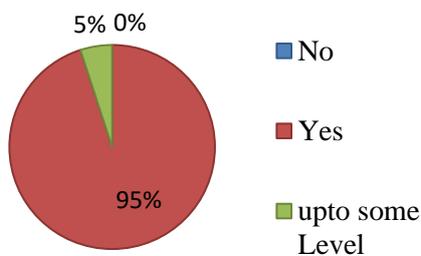

**Figure 8: Responses for navigation issues in website 2**

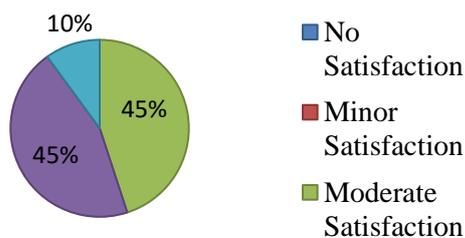

**Figure 9 Responses for Symantec annotations of the page elements**

According to results, 10% of them recorded a very high satisfaction level of using the feature, 45% responded that they are highly satisfied with the elements, and another 45% of the participants revealed that they are at a moderate satisfaction level. The critical response is no one has selected the feature with no or little satisfaction. It indicated that all participants are satisfied with the features. Therefore, this obstacle can overcome through semantic web and semantic annotations of the context of page elements.

The next problem identified was information overload due to excessive sequential reading, and the solution given includes voice method to get the information more quickly. The result is showing in the pie chart in Figure 10.

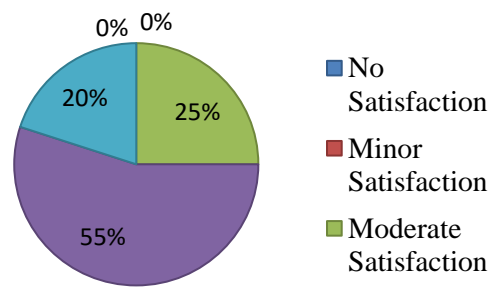

**Figure 10: Responses for concurrent speech methods**

According to the result, 20% of them chose the very high satisfaction level of using the feature, 55% responded that they are highly satisfied with the elements, and another 25% showed they are at a moderate satisfaction level. It indicated that all participants are satisfied with the features. Therefore, this obstacle can be avoided through concurrent speech methods. The next problem identified was content serialization. The pie chart in Figure 4.10 illustrates the satisfaction level of this feature by the visually impaired community.

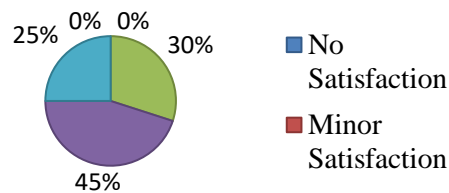

**Figure 11: Responses for content serialization**

According to the result, 25% of them reported a very high satisfaction level of using the feature, 45% responded that they are highly satisfied with the elements, and another 25% showed that they are at a moderate satisfaction level. It indicated that all participants are satisfied with the features. Therefore, it can conclude that content serialization is an essential factor to consider in web navigation.

The next section mainly focuses on how navigation through specialized keys could affect the quick access to required information. The results are shown in the pie chart in Figure 12.

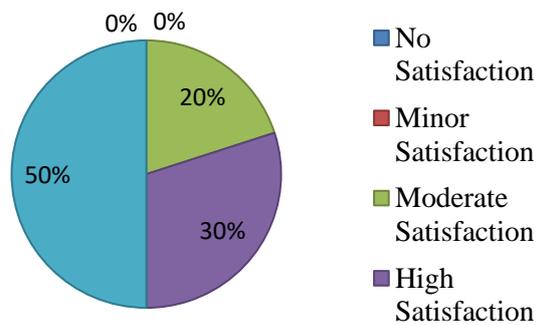

**Figure 12 Responses for specialised keys**

According to the result, 50% has a very high satisfaction level of using the feature, 30% responded they are highly satisfied with the elements, and another 20% showed that they are at a moderate satisfaction level. The critical response is none has selected this feature with no or little satisfaction. This indicates that all participants are satisfied with the elements. Hence, it is possible to overcome this obstacle by assigning special keys to the website to improve navigation.

Analysing the above results allows concluding that the visually impaired community needed the above-discussed features for an easy web navigation process.

## V. CONCLUSION AND RECOMENDATION

The Phase 1 covers the solutions concerning the access limited by impairment issues. Various proposed solutions were tested here; i.e., ensuring the website is keyboard-friendly, ensure that all contents are easily accessible and support with Symantec annotation, add Alt Text to All Images (Alternative Text for images), Use headers to structure the content correctly, design all forms for accessibility, Do not use tables for anything except tabular data, Enable resizable text that does not break a website, Avoid automatic media and navigation, Create content with accessibility in mind, and Use best colour combination for Visually Impaired Community. All these considerations were tested with the participation of the visually impaired individuals.

The findings revealed that the visually impaired community must receive these necessities to increase the improvement of accessibility, and thus, these are essential factors to overcome web accessibility issues of the community.

Phase II concerns on usability issues arising from lack of design considerations. This activity validates braille display (Braille Terminal) support and audio support designs. The result points out that audio support is a significant part of web accessibility due to vision difficulty. If the website provides a video without the help of audio, they are unable to acquire the correct information. Therefore, web designers and developers must consider pre-recorded videos with audio facilities. When considering braille support in web designing, it highlighted that the option does not directly impact visually impaired web users.

Phase III is concerned with the frustration of waiting until a visually impaired-friendly version is available; this issue can overcome by introducing a rating widget option. It was also evaluated by developing a website with the rating widget feature. Results concluded that this accessibility badge helps improve web accessibility among the visually impaired community. Hence, it is also considered as a significant feature to use in websites to overcome the web accessibility issues faced by the vision-impaired community.

Phase IV deals with the lack of communication issues in the requirement gathering and testing in website designing and development process. In this activity, the website is designed and developed with the involvement of using visually impaired users. It was further tested on how to relate its involvement with overcoming the disability digital divide. Results here detected a significant difference, and without the participation of this community in the web development process. It further revealed that visually impaired individuals must be involved in requirement gathering, during the user interface designing stage, and in testing accessibility. Also, testing only via accessibility compliance tools is not sufficient for this community to access a website.

Phase V is an in-depth study on solutions to overcome web navigation issues faced by the visually impaired community when they are accessed through screen reading software. The solution proposed to use the semantic web and semantic annotations of the context of Page elements, Content serialization, and navigation by special keyboard commands. According to the analysis of three web navigation aspects, respondents clearly stated that three of these aspects highly influence the effective web use and increase the satisfaction level of the websites accessing process. Hence, web developers must consider these three factors during website development, as they contribute to overcoming the web accessibility issues faced by the visually impaired community.

The findings revealed that the visually impaired community confront difficulties on web and mobile accessing. Therefore, the last recommendation is focused on the national level consideration to alteration of national level web guidelines that the web could follow. Therefore, a specific set of guidelines are recommended for web developers to apply in the web application design and development process.

Furthermore, this guidelines can be recommended as a benchmark in the national level policy development. This framework concentrates more specifically on solving the problems faced by the visually impaired community in Sri Lanka, and therefore, in national-level policy development to overcome digital divide process, it could be considered in border aspects to access the related problems and implement respective solutions.